\documentclass[prd,superscriptaddress,showpacs]{revtex4}
\usepackage{mathrsfs}
\usepackage{amssymb}

\newcommand{\D}{{\mathrm{D}}}

\newcommand{\ep}{\epsilon}
\newcommand{\be}{\begin{equation}}
\newcommand{\ee}{\end{equation}} 
\newcommand{\eei}{\end{equation}\indent\indent}
\newcommand{\bc}{\begin{center}}
\newcommand{\ec}{\end{center}}
\newcommand{\ber}{\begin{eqnarray}}
\newcommand{\eer}{\end{eqnarray}}
\newcommand{\ba}{\begin{array}}
\newcommand{\ea}{\end{array}}

\newcommand{\sfrac}[2]{{\textstyle{#1\over#2}}}
\def\case#1/#2{\textstyle\frac{#1}{#2} }
\newcommand{\bra}[1]{\left(#1\right)}
\newcommand{\bras}[1]{\left[#1\right]}
\newcommand{\brac}[1]{\left\{#1\right\}}
\newcommand{\curl}{{\mathsf{curl}\,}}
\newcommand{\di}{{\mathsf{div}\,}}
\newcommand{\reff}[1]{(\ref{#1})}
\begin{document}


\title{Primordial magnetic seed field amplification by gravitational waves}

\author{Gerold Betschart}
\email{gerold@phys.huji.ac.il} \affiliation{Department of
Mathematics and Applied Mathematics,
  University of Cape Town, 7701 Rondebosch, South Africa}

\author{Caroline Zunckel}
\affiliation{Department of Mathematics and Applied Mathematics,
  University of Cape Town, 7701 Rondebosch, South Africa}
\author{Peter K S Dunsby}
\affiliation{Department of Mathematics and Applied Mathematics,
  University of Cape Town, 7701 Rondebosch, South Africa}
\affiliation{South African Astronomical Observatory, Observatory
7925, Cape Town, South Africa}
\author{Mattias Marklund}
\affiliation{Department of Physics, Ume{\aa} University, SE-901 87
Ume{\aa}, Sweden}

\begin{abstract}

Using second-order gauge-invariant perturbation theory, a
self-consistent framework describing the non-linear coupling
between gravitational waves and a large-scale homogeneous magnetic
field is presented. It is shown how this coupling may be used to
amplify seed magnetic fields to strengths needed to support the
galactic dynamo. In situations where the gravitational wave
background is described by an `almost'
Friedmann-Lema{\^i}tre-Robertson-Walker (FLRW) cosmology we find
that the magnitude of the original magnetic field is amplified by
an amount proportional to the magnitude of the gravitational wave
induced shear anisotropy and the square of the field's initial
co-moving scale. We apply this mechanism to the case where the
seed field and gravitational wave background are produced during
inflation and find that the magnitude of the gravitational boost
depends significantly on the manner in which the estimate of the
shear anisotropy at the end of inflation is calculated. Assuming a
seed field of $10^{-34}$ $\rm{G}$ spanning  a comoving scale of
about $10\, \rm{kpc}$ today, the  shear anisotropy  at the end of
inflation must be at least as large as $10^{-40}$ in order to
obtain a generated magnetic field of the same order of magnitude
as the original seed. Moreover, contrasting the weak field
approximation to our gauge-invariant approach, we find that while
both methods agree in the limit of high conductivity, their
corresponding solutions are otherwise only compatible in the limit
of infinitely long-wavelength gravitational waves.
\end{abstract}


\maketitle

\section{Introduction}

The origin of cosmological magnetic fields that are prevalent
throughout galaxies clusters, disk and spiral galaxies and
high-redshift condensations has generated much debate in recent
years, with the majority of this work being focused on providing
mechanisms that generate these galactic fields on large scales
(see \cite{grasso,widrow} and references therein). The candidate
mechanisms are diverse, often depending on the required seed field
strengths. It has been suggested that the fields observed today
could be a result of the amplification of a relatively large seed
field through protogalactic collapse at the onset of structure
formation \cite{Roy2}. As the gas collapses to current measured
densities, the flux lines of the frozen-in cosmological magnetic
field get compressed, inducing adiabatic amplification. Another
popular mechanism, which requires a relatively weaker pre-existing
seed field, is amplification via the galactic dynamo by means of
parametric resonance \cite{Dynamo}. The combined effect of
differential rotation across the disk and the cyclonic turbulent
motions of the ionized gas is believed to lead to the exponential
amplification of a smaller primordial field until the
back-reaction of the plasma opposes further growth. Although the
dynamo mechanism is strongly supported by the close correlation
between the observed structure of the galactic fields and the
spiral pattern of galaxies, there is some argument over its
efficiency and hence the amount of amplification that can occur
through this process. The major problem with all of these
mechanisms is that they assume the presence of a pre-existing seed
field whose origin is still to be established. A further idea
relies on turbulence (disrupted flow) and shocks, which occur
during the stages of structure formation, inducing weaker magnetic
fields via battery-type mechanisms, which operate as a result of
large-scale misalignments of gradients in electron number density
and pressure (or temperature) \cite{davies}.

There have been numerous attempts to generate early,
pre-recombination, magnetic fields with strengths suitable to
support and maintain the dynamo  by exploiting the different
out-of-equilibrium epochs that are believed to have taken place
between the end of the inflationary era and decoupling
\cite{cheng}. These fields are facilitated by currents that arise
from local charge separation generated by vortical velocity fields
prevalent in the early plasma (cf. also \cite{bert}).

One problem with the above mechanisms is that they are casual in
nature so the scales over which the fields are coherent cannot
exceed the particle horizon during that epoch. Given that such
phase transitions took place at very early times, where the
comoving horizon size was small, tight constraints must be placed
on the coherence length of these magnetic fields. However, pre big
bang models based on string theory~\cite{string}, in which vacuum
fluctuations of the magnetic field are amplified by the dilaton
field, predict super-horizon fields.

Inflation has long been suggested as a solution to the causality
problem, since it naturally achieves correlations on superhorizon
scales, however adjustments to the standard inflationary models
need to be made since magnetic fields surviving this epoch are
small on account of the inability of vector fields to couple
gravitationally to the conformally flat metric resulting from the
exponentially fast expansion. A way around this obstacle is by
breaking the conformal invariance of electromagnetism since this
alters the way the underlying gauge fields couple to gravity.
There are many ways of doing this which explains the variety of
the proposed mechanisms in the literature \cite{conf}. Such
inflationary scenarios have not been without critique, though
\cite{critique}.

It has also been proposed that inflation is followed by a period
of preheating in which the parametric resonance of the causal
oscillations of the inflaton field and the accompanying
perturbations can lead to amplification on super-horizon
scales~\cite{preheating}. Other authors have advocated the
breakdown of Lorentz invariance either in the context of string
theory and non-commutative varying speed of light theories, or due
to the dynamics of large extra dimensions~\cite{BM}. The success
of these proposals, however, is usually achieved at the expense of
simplicity.

In order for these proposed mechanisms to be viable, they must, in
addition, produce seed fields that satisfy the criteria for the
subsequent amplification processes to work.  To be a candidate
seed field for the galactic dynamo, the induced field must exceed
a minimum coherence scale in order to prevent the destabilization
of the dynamo action. The time scale over which the amplification
takes place also dictates a minimum field strength, for example in
the case of a dark-energy dominated Universe we obtain $B\sim
10^{-34}~\mathrm{G}$ on a coherence scale of $10\ \mathrm{kpc}$.
Davis {\it et al.} \cite{D} proposed an inflationary mechanism
that exploits the natural coupling between the Z-boson and the
gravitational background. Unfortunately, the fields produced only
just fall within dynamo limits in the case of a dark energy
dominated Universe. Recently, the production of a magnetic seed
field due to the rotational velocity of ions and electrons, caused
by the nonlinear evolution of primordial density perturbations in
the cosmic plasma during pre-recombination radiation and matter
eras, was investigated in~\cite{Matt} and a rms amplitude
$B\approx 10^{-23}(\lambda/\mathrm{Mpc})^{-2}\,\mathrm{G}$ at
recombination on comoving scales $\lambda\gtrsim 1 \,\mathrm{Mpc}$
was reported.

In this paper, we offer an alternative mechanism that looks at the
interaction of a \emph{pre-existing} field, such as the one
proposed by Davis {\it et al.}, with a gravitational wave (GW)
spectrum which accompanies most inflationary scenarios. This builds
on earlier work by Tsagas {\it et al.} \cite{GWamp} in which
this idea was first introduced within the weak field approximation.
Our aim is to investigate whether this interaction can produce a
sufficiently large amplification of a seed field present at the end
of inflation to meet the above mentioned requirements for the dynamo to
work.

The issue of how to deal with the coupling between gravitational
waves and the seed magnetic field is rather subtle. A commonly
used approximation in the literature is to assume that the
magnetic field is weak and that its contribution to the
energy-momentum tensor is such that it does not disturb the
isotropy of the FLRW background \cite{tsagas}. This is done by
assuming that the energy density of the magnetic field
$\tilde{B}_a$ is much less than the matter energy density: $\tilde
B^2\ll\mu$ and that its anisotropic pressure is negligible:
$\pi_{ab}
\equiv-\tilde{B}_{<a}\tilde{B}_{b>}\approx0$~\footnote{Here the
angle bracket represents the projected symmetric trace-free (PSTF)
part of any tensor: $A_{<ab>}\equiv
h^c{}_{(a}h^d{}_{b)}A_{cd}-\sfrac{1}{3}h_{ab}A^c{}_c$.}. The
problem with this approximation is that it is not gauge-invariant
in a strict mathematical sense, so one can therefore not guarantee
that, when calculating the magnetic field which arises through its
coupling with linear perturbations of FLRW (such as gravitational
waves), it leads to physically meaningful results. In order to
solve this problem we develop a self-consistent framework based on
second-order perturbation theory, employing the methods initiated
by recent work of Clarkson~\cite{Clarkson} and Clarkson {\it et
al.} \cite{clarkson}. Here the seed magnetic field is treated as a
\emph{on average} homogeneous linear perturbation of the
background FLRW model and couplings to gravitational degrees of
freedom that arise when perturbing the background are taken to be
second order in the perturbation theory. Adopting this approach
allows us to write Maxwell's equations in a way that makes them
manifestly gauge-invariant to second order with interaction terms
that clearly describe the modes induced by the gravity
wave-magnetic field interaction. The restriction to a homogeneous
seed field leads to simplification on the technical level but
still encapsulates the main features of the gravito-magnetic
interaction. The implementation  of an inhomogeneous seed is
reserved to a future article.

The results show that, in the presence of gravitational radiation,
the magnitude of the magnetic field is amplified proportionally to
the shear distortion caused by the propagating waves. Once the
amplification is saturated, the magnetic field dissipates
adiabatically as usual. The gravitational boost is also
proportional to the square of the field's original scale, which
suggests that the proposed mechanism could lead to significant
amplification in the case of large scale magnetic fields. Indeed,
when applied to fields of roughly $10^{-34}~\mathrm{G}$ spanning a
comoving scale of about $10~\mathrm{kpc}$ today (see for example
the fields produced in~\cite{D}) , the  mechanism leads to an
amplification of up to 13 orders of magnitude (depending on the
calculation of the shear distortion), bringing these
magnetic fields well within the galactic dynamo requirements,
without the need for extra amplification during reheating. We thus
qualitatively and quantitatively rediscover in a gauge-invariant
fashion  the main results reported in \cite{GWamp}.

In order to contrast the two different approaches in detail, we
compare our solutions with the corresponding solutions obtained
using the weak field approximation \cite{tsagas} and find that
while both methods agree in the limit of high conductivity, their
corresponding solutions are otherwise only compatible in the limit
of infinitely long-wavelength gravitational waves when merely the
dominant contribution is considered.

The units employed in this paper are $c=h=1$ and $\kappa=8\pi
G=1$, the exception being section \ref{app}, where natural units
are used.
\section{Perturbation scheme}

If we wish to study the interaction between gravitational waves
and a magnetic field in a cosmological setting, we immediately
face a second-order problem in perturbation theory because both
the magnetic field as well as GW are absent in the exact FLRW
background, and may thus be individually regarded as first order
perturbations. Using the 1+3 covariant approach \cite{covariant},
we therefore develop a two parameter expansion in two smallness
parameters: $\epsilon_{B}$ represents the magnitude of a
homogeneous magnetic field and $\epsilon_g$ represents the
magnitude of the GW. The magnitude of the interaction GW $\times$
magnetic field is of order $\mathcal O(\epsilon_{B}\epsilon_{g})$
as is the magnitude of the in such a manner generated
electromagnetic fields. However, at second-order level, only terms
of order $\mathcal O(\epsilon_{B}\epsilon_{g})$ are kept while
terms of order $\mathcal O(\epsilon_g^2)$ and $\mathcal
O(\epsilon_{B}^2)$ are discarded. In fact, when dealing with the
gravito-magnetic interaction, these discarded terms would always
appear multiplied by a first-order quantity and are thus
irrelevant for our considerations.

Whence, the perturbation spacetimes are divided up and denoted in
the following way:
\begin{itemize}
\item{$\mathcal B$ = Exact FLRW as background spacetime,
$\mathcal{O}(\epsilon^0)$;} \item{$\mathcal F_1$ = Exact FLRW
perturbed by a homogeneous magnetic field whose energy density and
curvature are  neglected, $\mathcal O(\epsilon_{B})$;}
\item{$\mathcal F_2$ = Exact FLRW with gravitational perturbations
$\mathcal O(\epsilon_{g})$;} \item{$\mathcal S$ = $\mathcal
F_1+\mathcal F_2$ allows for inclusion of interactions terms of
order $\mathcal O(\epsilon_{B}\epsilon_{g})$.}
\end{itemize}
We will generally refer to terms of order  $\mathcal
O(\epsilon_{B})$ and $\mathcal{O}(\epsilon_g)$ appearing in
$\mathcal F$ as `first-order' and to variables of mixed order
$\mathcal O(\epsilon_{B}\epsilon_{g})$ appearing in $\mathcal S$
as `second-order'.

It should be noticed that the absence of an electric field in
$\mathcal F_1$ does not necessarily imply that there is no
electric field at all but rather that the electric field is
\emph{perturbatively} smaller than the magnetic field. This is in
accordance with the standard assumption that the very early
Universe was a good conductor (see, for example,
\cite{Baym-Heiselberg} for an example of how this works). The
inclusion of an electric field in $\mathcal F_1$ is possible, in
principle, but would require to alter the perturbation scheme
because then interactions between gravitational waves and the
electric field needed to be taken into account as well. However, a
more realistic way of describing the interaction between
gravitational waves and electromagnetic fields should employ a
multifluid description \cite{bert}, which allows for modelling the
currents, but that is beyond the scope of the present paper.

Having outlaid the different stages we turn to review the
concomitant equations. We keep them as general as possible, which
will allow us to illuminate the effects of spatial geometry,
cosmological constant $\Lambda$ and equation of state for the
matter on the interaction. We limit ourselves to the irrotational
case, that is, we require the vorticity $\omega_{ab}$ to vanish
throughout.
\subsection{FLRW background}\label{sec:phase}

The FLRW models are characterized by a perfect fluid matter tensor
and the condition of everywhere-isotropy. Thus, relative to the
congruence of fundamental observers with 4-velocity $u^a$
$(u^au_a=-1)$, the kinematical variables have to be locally
isotropic, which implies the vanishing of the 4-acceleration $\dot
u_a \equiv u^b\,\nabla_b\, u_a$, shear $\sigma_{ab} \equiv
\D_{<a}\,u_{b>}$ and vorticity $\omega_{ab} \equiv
\D_{[a}\,u_{b]}$ $(0=\dot u_a=\sigma_{ab}=\omega_{ab})$.
Furthermore, the models have to be not only conformally flat, that
is, the electric and magnetic components of the Weyl tensor vanish
$(0=E_{ab}=H_{ab})$, but also spatially homogeneous implying the
vanishing of the spatial gradients of the energy density $\mu$,
the pressure $p$ and the expansion $\Theta \equiv \D_a\, u^a$
$(0=\D_a\,\mu=\D_a\,\Theta=\D_a\,p)$. As usual, the spatial
derivative $\D_a \equiv h_a^{~b}\,\nabla_b$ is obtained by
projection of the spacetime covariant derivative $\nabla_a$ onto
the 3-space (with metric $h_{ab} \equiv g_{ab}+u_au_b$) orthogonal
to the observer's worldline. As a consequence, the key background
equations are the energy conservation equation \be
\dot{\mu}+\Theta(\mu+p)=0\ , \label{energy}\ee the Raychaudhuri
equation \be
\dot{\Theta}=-\sfrac{1}{3}\Theta^2-\sfrac{1}{2}\bra{\mu+3p}+\Lambda\
, \label{Raychaudhuri}\ee and the Friedmann equation \be
\mu+\Lambda=\frac{1}{3}\Theta^2+\frac{3\,K}{a^2}\
,\label{Friedmann} \ee
 where the constant $K$ indicates the geometry of
the spatial sections.
\subsection{First-order perturbations}\label{sec:phase1}

\subsubsection{The homogeneous magnetic field $\tilde B_a$}
We assume the magnetic field $\tilde B_a$ to be spatially
homogeneous at first order $(\D_a\tilde B_b=0)$ and thus consider
the gradient of $\tilde B_a$ as well as the magnetic anisotropy
$\Pi_{ab}=-\tilde B_{< a}\tilde B_{b> }$ as being of second order.
We presuppose that such a field was produced by some primordial
process, which left a relic field on average homogeneous over a
typical coherence length. Since there are no electric fields or
charges in the $\mathcal F_1$ perturbation spacetime, the magnetic
induction equation takes the form \be
\beta_a\equiv\dot{\tilde B}_{<a>}+\sfrac{2}{3}\Theta\tilde B_{a}=0\
. \label{induction} \ee As a result, the magnetic field scales as
\be \tilde B_a=\tilde B^0_a\bra{\frac{a_0}{a}}^2\ ,\label{Bscale}
\ee where $a$ denotes the scale factor, e.g., $\Theta = 3\, \dot
a/a = 3 H$, where $H$ denotes the inverse Hubble length.
\subsubsection{Gravitational waves}
Gravitational waves are covariantly described via transverse parts
of the electric $(E_{ab})$ and magnetic $(H_{ab})$ Weyl
components, which are PSTF tensors \cite{GW}. The pure tensor
modes are transverse, obtained by switching off scalar and vector
modes $(0=\D_a\mu= \D_a \Theta=\D_ap=\omega_a=\dot u_a)$, which
results in the constraints \footnote{We use $\curl V_a \equiv
\epsilon_{abc}\,D^b\,V^c$ to denote the $\curl$ of a vector and
$\curl W_{ab} \equiv \epsilon_{cd<a}\,\D^c\,W_{b>}^{~~d}$ to
denote the covariant $\curl$ of a second-rank PSTF tensor, where
$\epsilon_{abc}$ is the volume element of the 3-space. Finally,
the covariant spatial Laplacian is $\D^2 \equiv \D^a\D_a$.}
\be
0=\D^a\sigma_{ab}=\D^aE_{ab}=\D^aH_{ab}=H_{ab}-\curl\sigma_{ab}\ .
\ee
The propagation equations for these tensor modes are simply
\ber
\dot{\sigma}_{<ab>}+\sfrac{2}{3}\Theta\sigma_{ab}&=&-E_{ab}\ ,
\label{eq:shear}\\ \dot{E}_{<ab>}+\Theta E_{ab}&=&\curl(\curl
\sigma_{ab})-\sfrac{1}{2}\bra{\mu+p}\sigma_{ab}\ , \label{eq:dotE}
\eer
together with the background equations for $\Theta$ and
$\mu$. Since every FOGI tensor satisfies the linearized identity
\be
\curl(\curl T_{ab})=-\D^2T_{ab}+\sfrac{3}{2}\D_{< a}\D^c
\, T_{b> c}+ \bra{\mu+\Lambda-\sfrac13\Theta^2}T_{ab}\
,\label{cci}
\ee
we see that the gravitational waves are completely determined by a
closed wave equation for the shear, namely
\be
\ddot{\sigma}_{ab}-\D^2 \sigma_{ab}
+\sfrac{5}{3}\Theta\dot{\sigma}_{ab}+\bra{\sfrac19\Theta^2+\sfrac{1}{6}\mu
-\sfrac{3}{2}p+\sfrac{5}{3}\Lambda } \sigma_{ab}=0\ .
\label{eq:ddotsigma}
\ee
\subsection{The interaction}\label{sec:phase2}
Maxwell's equations govern the interaction between GW and magnetic
fields. If we require charge neutrality and neglect currents as
well as the back-reaction of induced second-order magnetic fields
with the shear, we obtain
\ber
\dot{E}_{<a>}+\sfrac{2}{3}\Theta E_{a} &=& \curl B_a\ ,\label{eq:dotEa} \\
\dot{B}_{<a>}+\sfrac{2}{3}\Theta B_{a} &=&\sigma_{ab}\tilde B^b - \curl E_a\ ,
 \label{eq:dotB}\\
D^aE_{a}&=&0\ ,\\
D^a{B}_{a}&=&0\ .
\eer
Observe that the EM fields have to be divergence-free at all orders
due to neglecting vorticity effects. Moreover, the system is not
gauge-invariant because it contains a mixture of second-order $(E_a,
\curl E_a, \curl B_a)$ and first-order terms $(\sigma_{ab})$, while
$B_a$ now comprises the full magnetic field (the first-order
contribution plus the induced field). The situation we are
interested in is the interaction between the shear $\sigma_{ab}$ and
the first-order magnetic field, neglecting the back-reaction with
the induced magnetic field. How does one then disentangle the
different magnetic field perturbations in a consistent way?

In special relativity, the standard procedure would be to use a
power series expansion of the magnetic field,
\be
  B^a=\epsilon_{B} B^a_{1}+ \epsilon_g\epsilon_{B}
B^a_{2}+\mathcal O(\epsilon^2_g,\epsilon^2_{B})\ , \label{eq:expB}
\ee
where the first-order field $B^a_1$ satisfies the magnetic
induction equation \reff{induction}. Although insertion of
this expansion into the above system yields only second-order
terms, the procedure does not work in general relativity since the
commutation relations for the various differential operators (cf.
the appendix) can not be consistently satisfied. To illustrate this
important point clearly, we consider the commutation relation
between the (proper) time derivative and the spatial gradient
applied to the magnetic field. It is evident that the case where
the commutator relation is introduced after the expansion of
$B^a$, \be \bra{D^b B^a}^{\dot{}}_{\perp}=
\epsilon_g\epsilon_B\bra{D^b B^a}^{\dot{}}_{\perp}=
\epsilon_g\epsilon_B \bras{D^b \dot{B}^a_{2}-\sfrac{1}{3}\Theta
D^b B^a_{2}}\ , \ee does not agree with the case where the
linearized identity for $(D^a B^b)^{\dot{}}$ is substituted before
using the power series expansion (\ref{eq:expB}): \ber \bra{D^b
B^a}^{\dot{}}_{\perp}&=&D^b \dot{B}^a-\sfrac{1}{3}\Theta D^b
B^a+H^{bd}\epsilon_{dac}B^c+\sigma^d_{~c} D^c B_a\nonumber\\
&=&\epsilon_g\epsilon_B \bras{D^b \dot{B}^a_{2}-\sfrac{1}{3}\Theta
D^b B^a_{2}}+\epsilon_B H^b_{~d}\epsilon^{dac}B_{c}^{1}\ .
\eer
Here, $\perp$ denotes projection onto the fundamental observer's
rest space. This inconsistency can only be resolved if all
interaction terms are zero. It is via the commutation relations
that Weyl curvature is brought in through the back door which
couples to the magnetic field and thus affects the interaction. It
is this feature that renders the power series procedure faulty.

The difficulty arises because the magnetic field $B^a$ is not
gauge-invariant in $\mathcal S$ as it does not vanish in $\mathcal
F_1$. We therefore need to define a new second order
gauge-invariant (SOGI) variable which satisfactorily describes the
effects that we wish to investigate. However, a look at Maxwell's
equations above reveals that
$\beta_a\equiv\dot{B}_{<a>}+\sfrac{2}{3}\Theta B_a$ is the sought
SOGI variable which has to be used at second order instead of the
magnetic field $B_a$. We chose to describe the interaction in
terms of the variable $I_a \equiv \sigma_{ab}\tilde B^b$. Hence,
Maxwell's equations can be written in truly gauge-invariant terms
at second-order, namely
\ber
\dot{E}_{< a>}+\sfrac{2}{3}\Theta E_a &=&\curl B_a\ ,
\label{eq:Ea}\\
\beta_a+ \curl E_a &=&I_a\ . \label{eq:beta} \eer Observe that the
standard constraints $0=\D^aB_a=\D^aE_a$, which hold at all
orders, imply \be \D^a\beta_a=\D^aI_a= \sigma_{ab}\D^a\tilde B^b
=0\ ,\ee where the last equality is only true as long as spatial
gradients of $\tilde B^a$ are regarded as second-order. Clearly,
if the idealized assumption of infinite conductivity is made so
that all electric fields vanish, Maxwell'sequations reduce to
$\beta_a=I_a$. In this specific case, once the solution for $I_a$
is known, the (not gauge-invariant) generated magnetic field
measured by the fundamental observer can be obtained via a
standard integration of $\beta_a$. However, it is important to
stress that $\beta_a$ is the fundamental variable, whose deviation
from zero quantifies the evolution of the magnetic field at
second-order in a truly gauge-invariant manner.

\section{Wave equations for the main variables}
Having written the key Maxwell's equations as a system of
differential equations of purely SOGI variables, we now turn to
the derivation of wave equations for the electric and magnetic
fields. In doing this we make no assumptions about the spatial geometry
or the equation of state and also keep the cosmological constant; this
has the advantage of allowing us to draw some conclusions about how these
parameters influence the interaction between GW and magnetic fields.
In particular, it will turn out that neglecting the current in Maxwell's
equations and at the same time requiring a homogeneous magnetic field at
first-order level leads to consistent equations in spatially flat
models only.

\subsection{Wave equation for the interaction variable}
Let us first derive the wave equation for the interaction variable
$I_a=\sigma_{ab}\tilde B^b$. Even though the shear $\sigma_{ab}$
belongs to $\mathcal F_2$ and the magnetic field $\tilde B_a$ to
$\mathcal F_1$, the commutator relations do not lead to
ambiguities for $I_a$ since they manifest themselves only at
third-order in this case. In order to derive an evolution equation
for $I_a$, we need the auxiliary quantity $J_a \equiv E_{ab}\tilde
B^b$. Then, using equations \reff{induction}, \reff{eq:shear},
\reff{eq:dotE} and \reff{cci}, we arrive at the system
\ber \dot{I}_{< a>}+\sfrac{4}{3}\Theta I_a&=&-J_a\ ,\\
\dot{J}_{< a>}+\sfrac{5}{3}\Theta J_a&=&-\D^2I_a
+\bras{\sfrac12\bra{\mu-p}+\Lambda- \sfrac13\Theta^2}I_a\ ,\eer
where we employed that spatial gradients of the magnetic field are
second-order and thus $\D^2I_a=\D^2\bra{\sigma_{ab}\tilde
B^b}=\bra{\D^2\sigma_{ab}}\tilde B^b$. Eliminating the auxiliary
variable $J_a$ , the general closed wave equation for $I_a$ is
found to be
\be \ddot{I}_{< a >}-\D^2 I_a +3\Theta\dot{I}_{< a>}+
\bras{\sfrac{13}{9}\Theta^2-\sfrac{1}{6}\mu-
\sfrac{5}{2}p+\sfrac{7}{3}\Lambda}I_a=0\ . \label{eq:ddotI}
\ee
In the case of infinite conductivity, the solution to equation
\reff{eq:ddotI} instantly yields the solution of $\beta_a$, from
which the induced magnetic field measured by the fundamental
observer might be obtained by integration.
\subsection{Wave equation for the electric field}
To derive the wave equation for the induced electric field, we
first differentiate equation \reff{eq:Ea} and equate the result
with the second-order identity
\be \bra{\curl B_a}^{\dot{}}_{\perp}=-\Theta\,\curl B_a+\curl
\beta_a-H_{ab}\tilde B^b \ee
to obtain
\be
\ddot{E}_{< a>}+\sfrac{5}{3}\Theta\dot{E}_{< a >}
+\bras{\sfrac{4}{9}\Theta^2-\sfrac{1}{3}\bra{\mu+3p}
+\sfrac{2}{3}\Lambda}E_a=\curl\beta_a-H_{ab}\tilde B^b\ .
\label{eq:ddotE}
\ee
Secondly, using equation \reff{eq:beta} to substitute for $\curl
\beta_a$ above and the expansion
\be \curl\bra{\curl E_a} = -\D^2E_a-\bras{\sfrac29\Theta^2
-\sfrac23\bra{\mu+\Lambda}}E_a\ , \label{cce} \ee we find a forced
wave equation for the induced electric field, namely
\be \ddot{E}_{< a>}-\D^2 E_a
+\sfrac{5}{3}\Theta\dot{E}_{< a>}
+\bras{\sfrac{2}{9}\Theta^2+\sfrac13\bra{\mu-3p}+\sfrac43\Lambda}E_a=K_a\
, \label{eq:wE}
\ee
where the forcing term  $K_a \equiv \curl I_a -H_{ab}\tilde B^b
=\ep_{cd[a}\D\,\sigma^{~~c}_{b]}B^b$ has no divergence.
It is possible to show that the forcing term $K_a$, as well as
$\curl I_a$ and $H_{ab}\tilde B^b$, respectively, can be found
from the wave equation
\be
\ddot K_{< a>}-\D^2K_a +\sfrac{11}{3}\Theta \dot K_{< a>}
+\bras{\sfrac{22}{9}\Theta^2-\sfrac{1}{3}\bra{\mu+9p}
+\sfrac{8}{3}\Lambda} K_a=0\ . \ee For example, the wave equation
for $\curl I_a$ follows by taking the $\curl$ of equation
\reff{eq:ddotI} and using the expansion \reff{cce}, while the case
$H_{ab}\tilde B^b$ is similar  to the derivation of the wave
equation for the interaction term $I_a$.

It will be useful for later purposes to consider the electric
field's rotation. By taking the $\curl$ of equation \reff{eq:wE},
we immediately arrive at
\ber \bra{\curl E_a}^{\ddot{}}_{\perp}&-&\D^2\bra{\curl E_a}
+\sfrac{7}{3}\Theta\bra{\curl E_a}^{\dot{}}_{\perp} \nonumber \\
&+&\bras{\sfrac{7}{9}\Theta^2+\sfrac16\bra{\mu-9p}
+\sfrac53\Lambda}\curl E_a=\curl K_a\ . \label{eq:wcurlE}
\eer
Because $\curl\bra{H_{ab}\tilde
B^b}=-\D^2I_a+\bras{-\sfrac{5}{18}\Theta^2+\sfrac56\bra{\mu+\Lambda}}I_a$
holds, we note the interesting result
\be \curl K_a =
\bras{\sfrac{1}{18}\Theta^2-\sfrac16\bra{\mu+\Lambda}}I_a\
.\label{curlK}\ee That is, for a cosmological model with flat
spatial sections we have $\curl K_a = 0$ and, therefore, the
electric field's rotation is not induced by the interaction
between magnetic fields and GWs at second-order -- the generated
electric field is curl-free. As a consequence, the interaction
between magnetic field and GW produces the same magnetic field in
a case of a spatially flat Universe as in the limit of high
conductivity.

However, upon closer inspection of the forcing term $K_a$ in
equation \reff{eq:wE} one discovers that this term is actually
identically zero because of the identity \cite{vanElstThesis}
\be
0=\epsilon^{abc}\,V_b\bra{\D_d\,
A_c^{~d}}-2\,V_b\,\epsilon^{cd[a}\bra{\D_c\, A^{b]}_{~d}}\ , \ee
which holds for any vector $V_a$ and tensor $A_{ab}=A_{<ab>}$
perpendicular to the congruence $u_a$. Thus, equation \reff{curlK}
implies that our chosen perturbative scheme is only consistent if
the cosmological model is spatially flat (cf. also footnote
\ref{f1} below). In essence, we see that the requirement of having
a spatially homogeneous and thus a curl-free magnetic field at
first-order can only be achieved when the Universe is spatially flat.
Furthermore, the interaction between GW and a magnetic field generates
in this particular case no electric fields (at least to second order in
the perturbation scheme).
\subsection{The generated magnetic field}\label{sec:phase3}
We have already pointed out that for spatially flat models the
generated magnetic field follows directly from the interaction
variable since in this case we have $\beta_a = I_a$. For closed or
open models, however, a wave equation for $\beta_a$ is needed to
determine the induced magnetic field. The sought after equation
may be obtained by applying the constraint equation \reff{eq:beta} to
equation \reff{eq:wcurlE} and substituting for $\curl K_a$ via
equation \reff{curlK}. This leads to
\ber
\ddot{\beta}_{< a>}-\D^2\beta_a+\sfrac{7}{3} \Theta\dot{\beta}_{<
a>} +\bras{\sfrac{7}{9}\Theta^2
+\sfrac{1}{6}\bra{\mu-9p}+\sfrac{5}{3}\Lambda}\beta_a = \nonumber
\\ \ddot{I}_{< a>}-\D^2I_a
+\sfrac{7}{3}\Theta\dot{I}_{< a>}
+\bras{\sfrac{13}{18}\Theta^2+\sfrac{1}{3}\mu-\sfrac32p
+\sfrac{11}{6}\Lambda}I_a\ . \label{eq:ddotbeta} \eer Observe that
for models with flat spatial sections the lhs and rhs of the above
equation become identical -- in agreement with the comment
following equation \reff{eq:wcurlE}. A slight simplification is
achieved by employing equation \reff{eq:ddotI} yielding finally a
forced wave equation for $\beta_a$:
\ber
\ddot{\beta}_{< a>}-\D^2\beta_a+\sfrac{7}{3} \Theta\dot{\beta}_{<
a>} &+&\bras{\sfrac{7}{9}\Theta^2
+\sfrac{1}{6}\bra{\mu-9p}+\sfrac{5}{3}\Lambda}\beta_a =
\nonumber \\
&-&\sfrac{2}{3}\Theta\dot{I}_{< a>}-
\bras{\sfrac{13}{18}\Theta^2-\sfrac{1}{2}\bra{\mu+2p-\Lambda}}I_a\
. \label{eq:ddotb} \eer It is evident that the variable $I_a$ and
hence the gravitational waves source fluctuations in the magnetic
field variable $\beta_a$. Another way to derive equation
\reff{eq:ddotb} consists of differentiating Maxwell's equation
\reff{eq:beta} twice, using equation \reff{eq:Ea} to get rid off
the $\curl E_a$-term and applying the corresponding commutation
relations. This clearly demonstrates the consistency of our
approximation scheme.
\section{Solutions for flat Universes}\label{sec:phase4}
After having derived the fundamental equations governing the
interaction between GWs and magnetic fields as well as the
generated  electromagnetic fields, we turn to the task of solving
them. For the sake of simplicity, we investigate the solutions
only for spatially flat models with zero cosmological constant
$\Lambda$. We assume the matter to obey a barotropic equation of
state, $p=w\mu$, with constant barotropic index $w$.
\subsection{A useful time variable}\label{sec:phase5}
The background equations (\ref{energy}--\ref{Friedmann})
subject to the assumptions stated above imply the following
evolution equation for the scale factor:
\be
\frac{\ddot{a}}{a}+\frac{1}{2}\bra{1+3w}\bra{\frac{\dot{a}}{a}}^2=0\
.
\ee
By integrating once and choosing initial conditions such that
$\Theta_0\equiv\Theta(t_0)=3H_0$ for some arbitrary initial time
$t_0$ with $H=\dot a/a$ defining the Hubble radius, we obtain the following
solution for the expansion
\be
\frac13\Theta=\frac{\dot{a}}{a}=\frac{2}{3(1+w)(t-t_0)+2/H_0}\ .
\ee
Integrating once more, we find for the scale factor the solution
\be
a(t)=a_0\bras{\sfrac{3}{2}H_0\bra{1+w}\bra{t-t_0}+1}^{\sfrac{2}{3\bra{1+w}}}\
.
\ee
The introduction of a dimensionless time variable, $\tau$, defined
for $w \neq -1$ as
\be
\tau\equiv\sfrac{3}{2}H_0\bra{1+w}\bra{t-t_0}+1\ ,
\ee
will turn out to be extremely useful as it simplifies the
integration of almost all the equations considered later
irrespective of the barotropic index while taking the initial
conditions explicitly into account as well. For example, the scale
factor evolves simply as $a=a_0\,\tau^{2/\bra{3\bra{1+w}}}$ and the
Hubble radius is given by $H=H_0/\tau$. Moreover, $\tau=1$
corresponds to the initial time $t_0$. Note however that the $\tau$
variable cannot be used in the de Sitter limit $w\rightarrow -1$,
where the scale factor becomes $a(t)=a_0\exp(H_0(t-t_0))$.
\subsection{Generated magnetic field}
Since we are only considering Universes with \emph{flat} spatial
geometry, the induced magnetic field can be found by integrating
over $\beta_a$. To this end, it suffices to solve for the
interaction variable $I_a$. A standard harmonic decomposition
\cite{harmonics} is used to take care of the Laplacian operator.
We expand the shear $\sigma_{ab}=\sum_{k}\sigma^{(k)}Q^{(k)}_{ab}$
in pure tensor harmonics, where as usual $\dot Q^{(k)}_{<ab>}=0$
and $\D^2Q^{(k)}_{ab}=-(k^2/a^2)Q^{(k)}_{ab}$ hold. Moreover, each
gravitational wave  mode is associated with the physical wave
length $\lambda_{\mathrm{GW}}=2\pi a/k$. Since the magnetic field
in $\mathcal F_1$ obeys $\curl \tilde B_a=0$, it follows that
$\D^2\tilde B_a =-\curl(\curl \tilde B_a)=0$ and therefore that
the expansion of the magnetic field $ \tilde B_{a}=\sum_{n}\tilde
B^{(n)}Q^{(n)}_{a}$ in pure vector (solenoidal) harmonics reduces
to $\tilde B_a=\tilde B^{(0)}Q^{(0)}_{a}$, where $\tilde
B^{(0)}=\tilde B^0(a_0/a)^2$. This just means that the magnetic
field $\tilde B^a$ is spatially constant, e.g., in agreement with
the assumption of homogeneity~\footnote{In light of the commutator
relation \reff{cce}, which holds for $\tilde B^a$ in $\mathcal
F_1$, $\D_a\tilde B_b=0$ (which also leads to $\curl \tilde B_a=0$
in our approximation scheme) is only consistent for a spatially
flat Universe---in an open or closed Universe, a current is needed
to uphold the magnetic field's homogeneity.\label{f1}}. Of course,
the solenoidal harmonics also obey the relations $\dot
Q^{(n)}_{<a>}=0$ and $\D^2Q^{(n)}_{a}=-(n^2/a^2)Q^{(k)}_{a}$.
Perturbations in $\mathcal S$ are conveniently decomposed with the
vector harmonics \footnote{It should be kept in mind that all
above introduced harmonics are exclusively defined on the
background FLRW spacetime.} $V^{(\ell)}_a\equiv
Q_{ab}^{(k)}Q^b_{(n)}$, which are readily verified to fulfill the
standard requirements $\dot V^{(\ell)}_{<a>}=0$ and
$\D^2V^{(\ell)}_a=-(\ell^2/a^2)V^{(\ell)}_a$, where the wavenumber
$\ell$ satisfies $\ell^2=(k_a+n_a)(k^a+n^a)$. Because  the
magnetic field in $\mathcal F_1$ has got only the zero mode in our
investigation, the wavenumber $\ell$ coincides with the wavenumber
$k$ of the shear.

Using the unified time variable $\tau$ and the harmonics explained
above, we transform the wave equation \reff{eq:ddotI} for the
interaction variable $I_a$ into an ordinary differential equation:
\be
\frac94\bra{1+w}^2
I''_{(\ell)}+\frac{27\bra{1+w}}{2\tau}I'_{(\ell)}
+\bras{\bra{\frac{\ell}{a_0H_0}}^2\tau^{-\frac{4}{3\bra{1+w}}} +
\frac{25-15w}{2\tau^2}}I_{(\ell)}=0\ ,\label{inttau}
\ee
where a prime means differentiation with respect to $\tau$.
Initial conditions are chosen as follows:
\ber
I_{(\ell)}(t_0)&=&\sigma_{(k)}(t_0)\tilde B_{0}\ , \label{IC1} \\
I_{(\ell)}'(\tau=1)&=&\tilde
B_0\bras{\sigma'_{(k)}(1)-\frac{4}{3(1+w)}\sigma_{(k)}(1)}\ ;
\label{IC2}
\eer
here, $\tilde B_0$ is the initial amplitude of the first-order
magnetic field and
\be
\dot\sigma_{(k)}(t_0)=3/2H_0\bra{1+w}\sigma'_{(k)}(1)
\ee
was used. For every mode $k$ we have initially $\sigma(t_0)=\sigma_0$ and
$\sigma'(1)=\sigma'_0$.
\subsubsection{Infinite-wavelength limit} \label{longl}
In the infinite-wavelength limit $(\ell\rightarrow 0)$, the
solution of equation \reff{inttau} is easily found to be
\be
    I^{(0)}(\tau)=C_1\,\tau^{-\frac{10}{3\bra{1+w}}}+
    C_2\,\tau^{\frac{-5+3w}{3\bra{1+w}}}\ ,\label{long}
\ee
where $C_1$ and $C_2$ are constants of integration. If the initial
conditions \reff{IC1}--\reff{IC2} are chosen, the corresponding
integration constants are
\ber
C_1&=&\frac{\bra{-5+3w}I_{(\ell)}(1)-3\bra{1+w}I'_{(\ell)}(1)}{5+3w}\
,\\
C_2&=&\frac{10I_{(\ell)}(1)+3\bra{1+w}I'_{(\ell)}(1)}{5+3w}\ .
\eer
We remark that this solution is in agreement with the result
obtained by multiplying the first-order magnetic field
\reff{Bscale} with the infinite-wavelength solution of the shear
equation \reff{eq:ddotsigma}. Whence, the total magnetic field in
the presence of infinite-wavelength GWs is
\be
B^{(0)}(\tau)=\tilde
B_0\,\tau^{-\frac{4}{3\bra{1+w}}}\bras{1-\frac{C_1}{\tilde
B_0H_0}\frac{2}{3\bra{1-w}}\bra{
\tau^{\frac{-1+w}{3\bra{1+w}}}-1}+\frac{C_2}{\tilde
B_0H_0}\frac{1}{1+3w}\bra{\tau^{\frac{2+6w}{3\bra{1+w}}}-1}}\ ,
\label{ampl}
\ee
where $\tilde B_0$ is the magnitude of the first-order magnetic
field interacting with the GW at initial time $t_0$ and it is
required for physical reasons that the induced magnetic field
vanishes initially. We stress that the interaction always leads to
an amplification of the magnetic field for any physically
acceptable choice of equation of state because of the growing
contribution in the second line of equation \reff{ampl}.

Let us look at some important special cases. For the sake of
simplicity, we take $I'_{(\ell)}(1)=0$ for granted. In the
matter-dominated era, where the matter is accurately described as
dust, $w=0$ and $a=a_0\tau^{2/3}$, this yields for  the magnetic
field mode \be B^{(0)}_{\mathrm{Dust}}(a)=\tilde
B_0\,\bra{\frac{a_0}{a}}^2\bras{1+\frac23\frac{\sigma_0}{H_0}
\brac{\bra{\frac{a_0}{a}}^{3/2}-1}
+\frac{2\sigma_0}{H_0}\brac{\frac{a}{a_0}-1}}\ ,\label{longdust}
\ee whereas for a radiation-dominated era, where  $w=1/3$ and
$a=a_0\tau^{1/2}$,  the  magnetic field mode is \be
B^{(0)}_{\mathrm{Rad}}(a)=\tilde
B_0\,\bra{\frac{a_0}{a}}^2\bras{1+\frac23\frac{\sigma_0}{H_0}
\brac{\frac{a_0}{a}-1}
+\frac56\frac{\sigma_0}{H_0}\brac{\bra{\frac{a}{a_0}}^2-1}}\
.\label{longrad} \ee It follows that in the infinite-wavelength
limit the amplification depends mainly on the scale factor and the
magnitude of the initial GW distortion relative to the Hubble
parameter $(\sigma/H)_0$.
\subsubsection{General case with $\ell\neq 0$}
The general solution to the interaction equation \reff{inttau} is:
\be
I_{(\ell)}(\tau)=\tau^{\frac{-5+w}{2\bra{1+w}}}
\bras{D_1\,J_1\bra{\frac{3w+5}{2\bra{1+3w}},\frac{\ell}{a_0H_0}\frac{2}{1+3w}
\tau^{\frac{1+3w}{3\bra{1+w}}}}
+D_2\,J_2\bra{\frac{3w+5}{2\bra{1+3w}},\frac{\ell}{a_0H_0}\frac{2}{1+3w}
\tau^{\frac{1+3w}{3\bra{1+w}}}}}\ ,\label{intsolgen}
\ee
where $D_1$, $D_2$ are integration constants and $J_1$, $J_2$
denote Bessel functions  of the first and second kind,
respectively. Observe that in the limit of infinite wavelengths,
$\ell\rightarrow 0$, the solution \reff{long} is recovered. The
generated magnetic field relative to the observer moving with
4-velocity $u^a$ can be calculated from the solution
\reff{intsolgen} analytically for every barotropic parameter $w$.
We will state here only the total magnetic field solution in the
case of dust and radiation, respectively. For dust, where $w=0$
and $a=a_0\tau^{2/3}$, the full magnetic field is \be
B^{(\ell)}_{\mathrm{Dust}}(a)= \tilde
B_0\,\bra{\frac{a_0}{a}}^2\bras{1+\frac{3}{4\pi^2}
\bra{\frac{\lambda_{\mathrm{GW}}}{\lambda_{\mathrm{H}}}}^2_0
\bra{\frac{\sigma_0}{H_0}+\frac{\sigma'_0}{2H_0}}
+\mathcal{O}(a^{-1})}\ , \label{mfdust} \ee while for radiation,
where  $w=1/3$ and $a=a_0\tau^{1/2}$,  the total magnetic field
modes obey \be B^{(\ell)}_{\mathrm{Rad}}(a)=\tilde
B_0\,\bra{\frac{a_0}{a}}^2\bras{1+\frac{3}{4\pi^2}
\bra{\frac{\lambda_{\mathrm{GW}}}{\lambda_{\mathrm{H}}}}^2_0
\bra{\frac{\sigma_0}{H_0}+\frac{2\sigma'_0}{3H_0}}+\mathcal{O}(a^{-1})}\
. \label{mfrad} \ee Here, we introduced the gravitational
wavelength $\lambda_{\mathrm{GW}}=2\pi a/k$ and the Hubble length
$\lambda_{\mathrm{H}}=1/H$. The un-displayed remainders
$\mathcal{O}(a^{-1})$ in the expressions above contain oscillating
functions which decay at least as fast as the inverse scale factor
$a^{-1}$. Note that when the infinite-wavelength limit of the full
solutions above is taken, the findings \reff{longdust} and
\reff{longrad} are rediscovered. The results
\reff{mfdust}--\reff{mfrad} clearly show how the generated
magnetic field depends on the initial conditions and that the late
time behaviour is almost identical for both dust and radiation. It
should be noted that the interaction can only be effective if the
wavelength of the GW matches the size of the magnetic field
region, $\lambda_{\mathrm{GW}}\sim \lambda_{\rm\tilde B}$: in the
case of $\lambda_{\mathrm{GW}}\gg\lambda_{\rm\tilde B}$ the
magnetic field cannot be physically affected by the GW, while for
$\lambda_{\mathrm{GW}}\ll \lambda_{\rm\tilde B}$ the effect
becomes negligible due to its  quadratic dependence on
$\lambda_{\mathrm{GW}}$. If we divide the findings
\reff{mfdust}--\reff{mfrad} through the energy density of the
background radiation, the dominant contribution can be summarized
as follows, \be \frac{B}{\mu^{1/2}_{\gamma}} \simeq
 \bras{1+\frac{1}{10}
\bra{\frac{\lambda_{\rm\tilde B}}{\lambda_{\mathrm{H}}}}^2_0
\bra{\frac{\sigma}{H}}_0}\bra{\frac{\tilde
B}{\mu^{1/2}_{\gamma}}}_0\ , \label{summary} \ee where the
wavenumber indices have been suppressed and $\sigma'_0 =0$ was
assumed. At late times, a significant amplification of the
original magnetic field can be achieved for super-horizon
gravitational waves. Note that a result almost identical to
\reff{summary} was obtained in \cite{GWamp}, wherein the factor
$1/10$ is replaced by $10$ instead. However, our result holds for
\emph{any} finite gravitational wavelength,
$\lambda_{\mathrm{GW}}\sim \lambda_{\rm \tilde B}$, while the
result in \cite{GWamp} assumes $\lambda_{\rm H} \ll
\lambda_{\mathrm{GW}}$. Moreover, \cite{GWamp} used somewhat
contrived initial conditions leading to an abrupt amplification of
the field whereas we chose initial conditions such that there is
no generated magnetic field present when the interaction kicks in
at the end of inflation.

\section{Application} \label{app}
In order to estimate the amplification of the seed field due to the
interaction with GWs we reproduce the analysis presented in
\cite{GWamp} using the same parameter values. We find it convenient
to adopt natural units in this section.

Given that the evolution of the (spatially flat) Universe is
dominated by a dark energy component such as a cosmological
constant or quintessence, the minimum seed required for the dynamo
mechanism to work is of the order of $10^{-30}\, \mathrm{G}$ at
the time of completed galaxy formation and coherent on a scale at
least as large as the largest turbulent eddy, roughly $\sim 100~
\mathrm{pc}$ \cite{Davis-Lilley-Törnkvist}. Such a collapsed
magnetic field corresponds to a field $\tilde B$ of $\sim\,
10^{-34}\, \mathrm{G}$ with coherence length $\lambda_{\rm \tilde
B} \sim 10\, \mathrm{kpc}$ on a comoving scale if the field
remains frozen into the cosmic plasma from the epoch of radiation
decoupling to galaxy formation. Its field strength compared to the
energy density of the background radiation, $\mu_{\gamma}$, gives
rise to the ratio $\tilde B/\mu_{\gamma}^{1/2}\sim 10^{-29}$,
which stays constant as long as the magnetic flux is conserved and
the magnetic field is frozen into the cosmic medium.

During inflation, the Hubble parameter $H$ remains constant and is
taken to be $H\sim 10^{13}\,\mathrm{GeV}$ \cite{D}. The scale of
the magnetic field therefore implies $\lambda_{\rm \tilde
B}/\lambda_{\rm H} \sim 10^{20}$ at the end of inflation. A
general prediction of all inflationary scenarios is the production
of large scale gravitational waves whose energy density per
wavelength  is roughly \cite{GWamp}
\be
\mu_{\mathrm{GW}} \simeq m_{\mathrm{Pl}}^2\bra{
\frac{1}{\lambda_{\mathrm{GW}}}}^2\bra{\frac{H}{m_{\mathrm{Pl}}}}^2\
. \label{energyGW}
\ee
Here, $\lambda_{\mathrm{GW}}$ denotes the wavelength of the GW and
$m_{\mathrm{Pl}}$ the Planck mass (see, for example,
\cite{books}). The total energy density of the gravity waves
expressed in terms of the shear is (see footnote 4 in \cite{ts}
for a neat discussion)
\be
\mu_{\mathrm{GW}} =
\frac{m_{\mathrm{Pl}}^2}{16\pi}\,\sigma_{ab}\,\sigma^{ab}\ ,\label{energyGW2}
\ee
which implies an induced shear anisotropy \cite{GWamp}
\be
    \bra{\frac{\sigma}{H}}_0 \simeq \bra{
\frac{\lambda_{\rm
H}}{\lambda_{\mathrm{GW}}}}_0\bra{\frac{H}{m_{\mathrm{Pl}}}}\ ,
\label{shearaniso}
\ee
where the zero suffix indicates the end of the inflationary epoch.
Typical inflationary models predict $H/m_{\mathrm{Pl}}\sim 10^{-6}$,
which lies comfortably within the bound $H/m_{\mathrm{Pl}}\lesssim
10^{-5}$ stemming from the quadrupole anisotropy of the CMB.

The interaction of such a primordial magnetic field with GWs
produced by inflation leads to a substantial amplification of the
former. Resorting to our result \reff{summary} and applying
\reff{shearaniso}, we find for the magnetic field \cite{GWamp} \be
\frac{B}{\mu^{1/2}_{\gamma}} \simeq
 \bras{1+\frac{1}{10}
\bra{\frac{\lambda_{\rm \tilde B}}{\lambda_{\mathrm{H}}}}_0
\bra{\frac{H}{m_{\mathrm{Pl}}}}}\bra{\frac{\tilde
B}{\mu^{1/2}_{\gamma}}}_0\ . \label{Bampli} \ee Substituting
$(\lambda_{\rm \tilde B}/\lambda_{\rm H})_0 \sim 10^{20}$ and
$H/m_{\mathrm{Pl}}\sim 10^{-6}$ into the above expression, we
obtain that GWs amplify the original magnetic field as much as 13
orders of magnitude. This mechanism thus brings an inflationary
seed such as in \cite{D} up to $\sim 10^{-21}\, \mathrm{G}$, which
is comfortably within the requirements of the galactic dynamo
mechanism \cite{Davis-Lilley-Törnkvist}. In Universes with zero
cosmological constant, the minimum seed for the dynamo has to be
raised from $\sim 10^{-30}\,\mathrm{G}$ to $\sim
10^{-23}\,\mathrm{G}$ \cite{Kulsrud}.\\
\\
However, the use of \reff{energyGW} and \reff{energyGW2} to find
the shear anisotropy is problematic in the sense that
\reff{energyGW} holds strictly speaking only on scales up to the
horizon; the energy stored in superhorizon modes cannot be
measured by local observers. Once the initially superhorizon GWs
re-enter the observer's horizon, they contribute to the measured
energy density. Therefore, the correct procedure is to use the
value of the shear anisotropy at horizon crossing,
$\bra{\sigma/H}_{HC}$, and scale that value back to the end of
inflation using its evolution equation. During the radiation
dominated era, the gravitational wave length varies with the scale
factor (i.e. $\lambda_{\rm GW}\sim \tau^{1/2}$), while the horizon
scales as the inverse of the square of the scale factor (i.e.
$\lambda_{\rm H}=H^{-1}\sim \tau$). This gives
\be
\bra{\frac{\lambda_{\mathrm{GW}}}{\lambda_{\mathrm{H}}}}=
\bra{\frac{\lambda_{\mathrm{GW}}}{\lambda_{\mathrm{H}}}}_0\tau^{-1/2}\
,
\ee
where $\tau=1$ corresponds to the end of inflation after
reheating. At the point in time where the gravitational wave
crosses back into the horizon, its physical wavelength equals the
Hubble scale $\lambda_{\rm H}$, that is
$\bra{\lambda_{\mathrm{GW}}/\lambda_{\mathrm{H}}}_{\mathrm{HC}}=1$.
Substituting for the values from above, we obtain the crossing
time $\tau_{\mathrm{HC}}=10^{40}$, which hitherto leads to the
Hubble parameter $H_{\mathrm{HC}}\sim 10^{-27}\mathrm{GeV}$ at
horizon crossing. During the radiation era, the temperature is
proportional to the square of the Hubble parameter \cite{books},
$T\sim\sqrt{m_{\mathrm{PL}}\,H/10}$, yielding a temperature of
$T_{\mathrm{HC}}\sim10^{-5}\,\mathrm{GeV}$ at horizon crossing.
Since $1~\mathrm{GeV}\sim 10^{13}~\mathrm{K}$, the actual
temperature is $\sim10^{8}~\mathrm{K}$, confirming that the
superhorizon mode under consideration indeed crosses back into the
horizon during the radiation-dominated epoch. Whence, combining
\reff{energyGW} and \reff{energyGW2} gives a shear anisotropy
\be
   \Sigma_{\mathrm{HC}} \equiv \bra{\frac{\sigma}{H}}_{\mathrm{HC}}
    \sim\bra{\frac{H}{m_{\mathrm{Pl}}}}_0
    \sim 10^{-6} \label{shearaniso2}
\ee
at the time of horizon crossing.

Assuming a flat model with no cosmological constant, the evolution
of the shear anisotropy, $\Sigma=\sigma/H$, can be obtained by
solving for the shear modes $\sigma_{(k)}$ from equation
\reff{eq:ddotsigma} and noting that $H\sim \tau^{-1}$. The result
is
\be \label{rs}
\Sigma_{(k)}(\tau)=\tau^{\frac{-1+9w}{6\bra{1+w}}}
\bras{A\,J_1\bra{\frac{3w+5}{2\bra{1+3w}},\frac{k}{a_0H_0}\frac{2}{1+3w}
\tau^{\frac{1+3w}{3\bra{1+w}}}}
+B\,J_2\bra{\frac{3w+5}{2\bra{1+3w}},\frac{k}{a_0H_0}\frac{2}{1+3w}
\tau^{\frac{1+3w}{3\bra{1+w}}}}}\ ,
\ee
where $A$, $B$ are constants of integrations and $J_1$, $J_2$
denote Bessel functions of the first and second kind,
respectively. Since we are only interested in the growing mode, we
may set $B=0$ eliminating the decaying mode contribution.
Specializing to the case of radiation, $w=1/3$,  remembering that
horizon crossing for the mode under consideration (for which
$k/(a_0 H_0)=2\pi \bra{\lambda_{\rm H}/\lambda_{\rm GW}}_0\sim
2\pi\times 10^{-20}$ holds) happens at
$\tau_{\mathrm{HC}}=10^{40}$ and using the estimate
\reff{shearaniso2} for the shear anisotropy at horizon crossing,
one determines the remaining constant to be
$|A|\sim10^{-16}\,\pi$. Hence, at the end of inflation $(\tau=1)$,
one finally obtains for the sought shear anisotropy
\be
\Sigma_{0}=\bra{\frac{\sigma}{H}}_0\sim 10^{-45}\ ,
\ee
where the approximation $J_1(\nu,x) \sim x^{\nu}$ for small
arguments $x\ll 1$ of the Bessel function of the first kind has
been employed. This is remarkably close to the result one would
obtain by simply using the growing mode solution of \reff{rs} in
the limit $k/(a_0H_0)\ll1$, that is $\Sigma_{(k)}=\Sigma_0\tau$ in
the case of radiation, which gives $\Sigma_0\sim10^{-46}$. If the
above value for the shear anisotropy is used in \reff{summary},
the gravito-magnetic amplification is completely negligible:
\be
\frac{1}{10} \bra{\frac{\lambda_{\rm\tilde
B}}{\lambda_{\mathrm{H}}}}^2_0 \bra{\frac{\sigma}{H}}_0\sim
10^{-6}\ .
\ee
\\
\\We stress that the efficiency of the mechanism depends crucially
on the ratio between the coherence length $\lambda_{\rm\tilde B}$
of the initial magnetic field and the initial size of the horizon
$\lambda_{\rm H}$. This ratio, however, disappears when the
infinite-wavelength limit is taken (see section \reff{longl}).
Even though the solutions \reff{longdust}, \reff{longrad} show a
growth proportional (quadratic) to the scale factor, the factor of
proportionality $(\sigma/H)_0$ $(\sim 10^{-26}$ or $\sim 10^{-45}$
in our first and second examples, respectively) is far too small
in order to achieve an effective amplification. It follows that
the interaction between GWs and on average homogeneous magnetic
fields is completely negligible in the limit of infinitely
long-wavelength gravity waves.
\section{Comparison and discussion}
The interaction between GWs and magnetic fields in the
cosmological setting has recently been investigated in
\cite{GWamp}, where the weak field approximation \cite{tsagas} was
used. Here one allows for a weak magnetic test field $\tilde B_a$
in the background, whose energy density, anisotropic stress and
spatial dependence have negligible impact on the background
dynamics: $\tilde B^2 \ll \mu$ and $\pi_{ab}=-\tilde B_{<a}\tilde
B_{b>}\simeq 0 \simeq \D_a\tilde B_b$ to zero order. In order to
isolate linear tensor perturbations, it is necessary to impose
$\D_a\tilde B^2=0=\ep_{abc}\,\tilde B^b\,\curl \tilde B^c$ in
addition to the standard constraints $\omega_a=0=\D_a\mu=\D_a p$
associated with pure perfect fluid cosmologies. In the weak field
approximation, the main equations governing the induced magnetic
field arising from the interaction between  a weak background
magnetic field $\tilde B^a$ and GWs were derived in \cite{GWamp}
for the case of a spatially flat Universe with vanishing
cosmological constant $\Lambda$ and a barotropic equation of state
$p=w\mu$:
\be
\ddot B_{(\ell)} + \frac53\Theta\dot B_{(\ell)}
+\bras{\frac13\bra{1-w}\Theta^2+\frac{\ell^2}{a^2}}B_{(\ell)}=
2\bra{\dot\sigma_{(k)}+\frac23\Theta\sigma_{(k)}}\tilde
B_0^{(n)}\bra{\frac{a_0}{a}}^2\ ,\label{Bindweak}
\ee
where the GWs are determined by the shear wave equation
\be \label{shearweak}
\ddot\sigma_{(k)}+\frac53\Theta\dot \sigma_{(k)}
+\bras{\frac16\bra{1-3w}\Theta^2+\frac{k^2}{a^2}}\sigma_{(k)}=0\ .
\ee
Here, the shear is harmonically decomposed as
$\sigma_{ab}=\sigma_{(k)}Q^{(k)}_{ab}$, while for the induced
magnetic field $B_a^{(\ell)}=B_{(\ell)}V_a^{(\ell)}$ with
$V_a^{(\ell)}=Q_{ab}^{(k)}Q_{(n)}^b$ was adopted. The background
magnetic field evolves as $\tilde B_a=\tilde B^0_a(a_0/a)^2$ and
$\tilde B^0_a=\tilde B^0_{(n)}Q_a^{(n)}$ is assumed.

We want to compare our results with the corresponding ones in the
weak field approximation. For simplicity, we restrict ourselves
here to the case of dust. As pointed out above, the only allowed
magnetic wavenumber for the interacting magnetic field is $n=0$,
when $\D_a\tilde B_b=0$, which leads to $\ell=k$. The published
solution for the generated magnetic field in the weak field
approximation, e.g., equation (21) in \cite{GWamp}, however, is
not applicable in the limit $n\rightarrow 0$. This can be traced
back to the choice for the initial conditions for the generated
magnetic field made by the authors of \cite{GWamp} when solving
equations \reff{Bindweak}--\reff{shearweak}, see equation (19) in
\cite{GWamp}.

In what follows below, we solve equations
\reff{Bindweak}--\reff{shearweak} again, including the full
solution for the shear instead of merely keeping the dominant part
as done in \cite{GWamp}. We specify the initial conditions by
choosing for every mode $k$ of the shear
$\sigma_{(k)}(a_0)=\sigma_0$, $\dot\sigma_{(k)}(a_0)=0$ and for
every mode $\ell=k$ of the generated magnetic field
$B_{(\ell)}(a_0)=0=\dot B_{(\ell)}(a_0)$. Note that this choice of
initial conditions differs from that in \cite{GWamp} but agrees
with our choice made in section IV. The solution, including the
background field, for an arbitrary wavenumber $k$ of the shear has
the structure \be B^{(\ell)}_{\mathrm{Dust}}(a) = \tilde
B_0\bra{\frac{a_0}{a}}^2\bras{1+\frac{\sigma_0}{H_0}f\bra{\sqrt a;
k }+\mathcal O\bra{a^{-\frac12}}}\ ,\label{mfdustweak} \ee where
the function $f\bra{\sqrt a; k }$ is built  of several oscillatory
terms with amplitude
$(\lambda_{\mathrm{GW}}/\lambda_{\mathrm{H}})_0^2$ at most and the
un-displayed part falls of at least as fast as $a^{-1/2}$. If this
is compared with our result \reff{mfdust}, one observes that it
differs by having another time behaviour. More strikingly,
however, is that now the term $f\bra{\sqrt a; k }$  not only
amplifies the seed field but also grows like $\sqrt a$ in the long
wavelength limit ($k/a_0H_0\ll 1$). This is in clear contrast to
the gauge-invariant result \reff{mfdust}, where the seed undergoes
amplification but then still decays adiabatically like $a^{-2}$.
On the other hand, in the infinite-wavelength limit (
$k\rightarrow 0$), the exact full solution is now
\be
B^{(0)}_{\mathrm{Dust}}(a)=\tilde
B_0\,\bra{\frac{a_0}{a}}^2\bras{1+\frac{\sigma_0}{H_0}
\brac{\frac{20}{3}-14\bra{\frac{a}{a_0}}^{1/2}
+\frac{36}{5}\bra{\frac{a}{a_0}}+\frac{2}{15}\bra{\frac{a_0}{a}}^{3/2}}}\
.\label{longdustweak}
\ee
Again, we obtain a solution whose time behaviour differs from that
found in \reff{longdust}. However, the weak field solutions agree
with our presented solutions in the infinite-wavelength limit when
only the dominant part of the solutions is considered, at least in
the examples considered above. The reason why the solutions
obtained within the weak field approximation are in general not
equivalent to the solutions found using the gauge-invariant
approach developed in this paper results from the
non-gauge-invariance of the weak field approximation, where the
magnetic field $\tilde B_a$ interacting with the GW is treated as
a weak background field. However, gauge-invariance requires
$\tilde B_a$ to vanish exactly in the FLRW background. We remind
the reader once more that our procedure solves firstly for the
gauge-invariant variable $\beta_a=\dot B_{<a>} +\sfrac23\Theta
B_a$, from which the magnetic field $B_a$ measured in the frame of
reference of $u^a$ can then subsequently be found.

A further important remark concerns the issue of conductivity. We
have seen earlier that, within our assumptions and for spatially
flat Universes, the gravito-magnetic interaction leads to an
induced magnetic field which is independent of the conductivity of
the cosmic medium. This is due to the fact that the interaction does
not generate rotational electric field modes which might affect the
magnetic field. In the weak field approximation, however, the situation is
completely different. If one assumes that the conductivity of the
cosmic medium that high so that electric fields are quickly
dissipated away, yielding a curl-free induced magnetic field, then
equation \reff{Bindweak} no longer applies and one simply has to
use
\be
\dot{B}_{(\ell)} + \frac{2}{3}\Theta B_{(\ell)} =
\sigma_{(k)}\tilde B_0^{(n)}\bra{\frac{a_0}{a}}^2
\ee
instead, while the equation for the shear \reff{shearweak} is
unaltered. This means that the weak field approximation produces
the same result as our gauge-invariant perturbation approach in
the high conductivity limit, and for that case only. It is therefore
evident that in the weak field approximation the conductivity of the
cosmic medium has a crucial bearing on the generated magnetic
field, in stark contrast to the result of our gauge-invariant
approach (see also \cite{Marklund-Clarkson}).

\section{Conclusion}
In this paper we have investigated the properties of magnetic
fields in the presence of cosmological gravitational waves, using
a two parameter perturbation scheme. Using proper second-order
gauge-invariant variables (SOGI), we were able to obtain results
in terms of clearly defined quantities, with no ambiguity
concerning the physical validity of the variables. The full set of
equations determining the evolution of the gravitational waves and
the generated electromagnetic fields was presented, and the
integration shows an amplification of the induced magnetic field
due to the interaction of a `background' magnetic field with
gravitational waves. The magnitude of the original magnetic field
is amplified by an amount proportional to the magnitude of the
gravitational wave induced shear anisotropy and the square of the
field's initial co-moving scale Once the amplification saturates,
the magnetic field dissipates adiabatically as usual. The results
were discussed in different fluid regimes, in particular dust and
radiation, and it was established that the dominant contribution
to the magnetic field is the same in both fluid regimes. We find
that the magnitude of the gravitational  boost depends
significantly on the manner in which the estimate of the shear
anisotropy at the end of inflation is calculated. For a seed field
of $10^{-34}$ $\rm{G}$ spanning  a comoving scale of about $10\,
\rm{kpc}$ today, the  shear anisotropy  at the end of inflation
(during which we assume $H~\sim 10^{13}\,\rm{GeV}$) should be
larger than $10^{-40}$ for any noticeable  amplification of the
seed field to arise at all.

Moreover, we further recalculated the induced magnetic field
employing the weak-field approximation, thereby extending previous
results in \cite{GWamp}, and compared the solutions with ours
derived in a gauge-invariant manner using SOGI variables. It was
found that there is a significant difference in the growth
behaviour of the magnetic field when SOGI variables are used as
compared to the case of a weak-field approximation scheme. While
the two methods agree in the limit of high conductivity, they seem
to be compatible otherwise only in the limit of infinitely
long-wavelength gravitational waves when the dominant part of the
solution is considered.

\acknowledgments

The authors are indebted to Dr. C.G.\ Tsagas for valuable comments
and discussions. This research was supported by a Sida/NRF grant.

\appendix*

\section{Commutation relations}
Here we present various commutator relations which have been used
in the text. The relations are given up to second
order in our perturbation scheme. The vanishing of vorticity,
$\omega_{ab}=0$, is assumed throughout in conjunction with the
constraints $\D_a\mu=\D_a p=0$ which isolate the pure tensor
modes. All appearing tensors are PSTF, $S_{ab}=S_{<ab>}$, and all
vectors $V_a,\,W_a$ are purely spatial.

Commutators for first-order vectors $V_a$:
\ber
\bra{\D_a V_b}^{\dot{}}_{\perp} &=& \D_a
\dot{V}_b-\sfrac{1}{3}\Theta
\D_a V_b -\sigma_a^{~c}\D_cV_b+H_a^{~d}\,\epsilon_{dbc}\,V^c \\
\bra{\curl V_a}^{\dot{}}_{\perp} &=&
\curl\dot{V}_a-\sfrac{1}{3}\Theta\,
\curl V_a -\epsilon_{abc}\,\sigma^{bd}\,\D_d\, V^c -H_{ab}\,V^b \\
\D_{[a}\,\D_{b]} V_c &=&
\bras{\sfrac{1}{9}\Theta^2-\sfrac{1}{3}\bra{\mu+\Lambda}}V_{[a}\,h_{b]c}+
\bra{\sfrac{1}{3}\Theta\,\sigma_{c[a}-E_{c[a}}V_{b]}
\nonumber\\&&+h_{c[a}\bra{E_{b]d}-\sfrac{1}{3}\Theta\,\sigma_{b]d}}
V^d
\eer
Commutators for first-order tensors $S_{ab}$:
\ber
\bra{\D_a S_{bc}}^{\dot{}}_{\perp} &=& \D_a
\dot{S}_{bc}-\sfrac{1}{3}\Theta \D_a S_{bc}-\sigma_a^{~d} \D_d
S_{bc}+2H_a^{~d}\,\epsilon_{de(b}\,S_{c)}^{~~e}
\\
\bra{\D^b S_{ab}}^{\dot{}}_{\perp} &=& \D^b
\dot{S}_{ab}-\sfrac{1}{3}\Theta \D^b S_{ab}-\sigma^{bc}\D_c
S_{ab}+\epsilon_{abc}\,H^b_{~d}\, S^{cd}
\\
 \bra{\curl S_{ab}}^{\dot{}}_{\perp}&=&\curl
\dot{S}_{ab}-\sfrac{1}{3}\Theta\, \curl S_{ab} -\sigma_e^{~c}\,
\epsilon_{cd(a}\,D^e S_{b)}^{~~d} +3 H_{c<a}S_{b>}^{~~c}
\\
\curl\curl S_{ab}&=&-D^2 S_{ab}+\bra{\mu+\Lambda
-\sfrac{1}{3}\Theta^2}S_{ab} +\sfrac{3}{2}D_{<a}D^c
S_{b>c}\nonumber\\
&&+3 S_{c<a}\bra{E_{b>}^{~~c}-\sfrac{1}{3}\Theta\sigma_{b>}^{~~c}}
\eer
Commutators for second-order vectors $W_a$:
\ber
\bra{\D_a W_b}^{\dot{}}_{\perp}&=&\D_a
\dot{W}_b-\sfrac{1}{3}\Theta\, \D_a W_b
\\
\D_{[a}\,\D_{b]} W_c &=&
\bras{\sfrac{1}{9}\Theta^2-\sfrac{1}{3}\bra{\mu+\Lambda}}W_{[a}\,h_{b]c}
\\
\curl\curl W_a &=& -\D^2W_a +\D_a \bra{\di W}
+\sfrac23\bra{\mu+\Lambda -\sfrac{1}{3}\Theta^2}W_{a}
\eer



\begin{thebibliography}{99}

\bibitem{grasso}
D. Grasso and H.R. Rubinstein, Phys. Rep. {\bf 348}, 163 (2001).

\bibitem{widrow}
L.M. Widrow, Rev. Mod. Pys. {\bf 74}, 775 (2002).

\bibitem{Roy2}
R. Maartens, Pramana J. Phys. \textbf{55}, 575 (2000).

\bibitem{Dynamo}
E.N. Parker, {\it Cosmological Magnetic Fields} (Clarendon,
Oxford, 1979);\\
Y.B. Zeldovich, A.A. Ruzmaikin, and D.D. Sokoloff, {\em
Magnetic Fields in Astrophysics} (McGraw-Hill, New York, 1983);\\
P.P. Kronberg, Rep. Prog. Phys. {\bf 57}, 325 (1994);\\
R. Beck, A. Brandenburg, D. Moss, A.A. Shukurov, and D.D.
Sokoloff,
 Annu. Rev. Astron. Astrophys. {\bf 34}, 155 (1996).

\bibitem{davies}
H. Lesch and M. Chiba, Astron. Astrophys. {\bf 297},
305 (1995);\\
R.M. Kulsrud, R. Chen, J.P. Ostriker, and D. Ryu, Ap. J. {\bf 480}, 481 (1997);\\
G. Davies and L.M. Widrow, Ap. J. {\bf 395}, 34 (1999).

\bibitem{cheng}
C.J. Hogan, Phys. Rev. Lett. {\bf 51}, 1488 (1983);\\
B. Cheng and A. Olinto, Phys. Rev. D {\bf 50}, 2421 (1994);\\
G.Sigl, A. Olinto, and K. Jedamzik, Phys. Rev. D {\bf 55}, 4582 (1997);\\
D.T. Son, Phys. Rev. D {\bf 59}, 094019 (1999);\\
R. Brustein and D.H. Oaknin, Phys. Rev. D {\bf 60}, 023508 (1999);\\
D. Boyanovski, H.J. de Vega, and M. Simionato, Phys. Rev. D {\bf
67}, 123505 (2003).

\bibitem{bert}
M. Marklund, P.K.S. Dunsby, G. Betschart, M. Servin, and C.G.
Tsagas, Class. Quantum Grav.  {\bf 20}, 1823 (2003);\\
G. Betschart, P.K.S. Dunsby, and M. Marklund, Class. Quant. Grav.
{\bf 21}, 2115 (2004).

\bibitem{string}
M. Gasperini, M. Giovannini, and G. Veneziano, Phys. Rev.
Lett. {\bf 75}, 3796 (1995);\\
D. Lemoine and M. Lemoine, Phys. Rev. D {\bf 52}, 1955 (1995).

\bibitem{conf}
A.D. Dolgov, Sov. Phys. JETP {\bf 54}, 223 (1981);\\
M.S. Turner and L.M. Widrow, Phys. Rev. D {\bf 30}, 2743 (1988);\\
W.D Garretson, G.B. Field, and S.M. Carroll, Phys. Rev. D {\bf 46}, 5346 (1992);\\
A.D. Dolgov, Phys. Rev. D {\bf 48}, 2499 (1993);\\
E. A. Calzetta, A. Kandus, F.D. Mazzitelli, and C.E.M. Wagner, Phys. Rev. D {\bf 57}, 7139 (1998);\\
A. Kandus, E.A. Calzetta, F.D. Mazzitelli, and C.E.M. Wagner,  Phys. Lett. B {\bf 472}, 287 (2000);\\
E. Calzetta and A. Kandus, Phys. Rev. D {\bf 65}, 063004 (2002);\\
A. Ashoorioon and R.B. Mann, preprint [arXive:gr-qc/0410053]
(2005).

\bibitem{critique}
 M. Giovannini and M. Shaposhnikov, Phys. Rev. D {\bf 62}, 103512 (2000).

\bibitem{preheating}
B.A. Bassett, C. Gordon, R. Maartens, and D.I. Kaiser, Phys. Rev. D {\bf 61}, 061302 (2000);\\
B.A. Bassett, G. Pollifrone, S. Tsujikawa, and F. Viniegra, Phys. Rev. D {\bf 62}, 103515 (2001);\\
F. Finelli and A. Grappuso, Phys. Lett. B {\bf 502}, 216 (2001).

\bibitem{BM}
O. Bertolami and D.F. Mota, Phys. Lett. B {\bf 455}, 96 (1999);\\
M. Giovannini, Phys. Rev. D {\bf 62}, 123505 (2000);\\
A. Mazumdar and M.M. Sheikh-Jabbari, Phys. Rev. Lett. {\bf 87},
011301 (2001).

\bibitem{D}
A.-C. Davis, K. Dimopoulos, T. Prokopec, and O.
T\"{o}rnkvist, Phys. Lett. B {\bf 501}, 165 (2001);\\
K. Dimopoulos, T. Prokopec, O. T\"{o}rnkvist, and A-C. Davis,
Phys. Rev. D {\bf 65}, 063505 (2002).

\bibitem{Matt}
S. Matarrese, S. Mollerach, A. Notari, and A. Riotto, Phys. Rev. D
{\bf 71},  043502 (2005).

\bibitem{GWamp}
C.G. Tsagas, P.K.S. Dunsby, and M. Marklund, Phys. Lett. B {\bf
561,} 17 (2003).

\bibitem{tsagas}
C.G. Tsagas and J.D. Barrow, Class. Quantum Grav. {\bf
14}, 2539 (1997);\\
C.G. Tsagas and J.D. Barrow, Class. Quantum Grav. {\bf
15}, 3523 (1998);\\
C.G. Tsagas and R. Maartens, Phys. Rev. D {\bf 61},
083519 (2000);\\
C.G. Tsagas and R. Maartens, Class. Quantum Grav. {\bf
17}, 2215 (2000);\\
S. Hobbs and P.K.S Dunsby, Phys. Rev. D {\bf 62}, 124007 (2000).

\bibitem{Clarkson}
C.A. Clarkson, Phys. Rev. D {\bf 70}, 103524 (2004).

\bibitem{clarkson}
C.A. Clarkson, M. Marklund, G. Betschart, and P.K.S. Dunsby, Ap.
J. {\bf 613}, 492 (2004).

\bibitem{covariant}
J. Ehlers, Abh. Mainz Akad. Wiss. Lit. (Math. Nat. Kl.)
{\bf 11}, 1 (1961);\\
S.W. Hawking, Ap. J. {\bf 145}, 544 (1966);\\
G.F.R. Ellis, ``Relativistic Cosmology" in {\em Carg\`{e}se
Lectures in Physics}, vol VI, ed. E. Schatzmann (Gordon and
Breach, 1973);\\
G. F. R. Ellis and H. van Elst,
  in \textit{Theoretical and Observational Cosmology}, pp.\ 1--116
  ed.\ Marc Lachi\`{e}ze-Rey (Kluwer, Dordrecht, 1999); gr-qc/9812046

\bibitem{Baym-Heiselberg}
    G. Baym and H. Heiselberg, Phys. Rev. D {\bf 56}, 5254 (1997).

\bibitem{GW}
P.K.S. Dunsby, B.A. Bassett, and G.F.R. Ellis, Class.
Quantum Grav. {\bf 14}, 1215 (1997); \\
A. Challinor, Class. Quantum Grav. {\bf 17}, 871 (2000).

\bibitem{vanElstThesis}
H. van Elst, Ph.D. thesis, (Queen Mary \& Westfield College,
London, 1996).

\bibitem{Davis-Lilley-Törnkvist}
A.-C. Davis, M. Lilley, and O. T\"ornkvist, Phys. Rev. D {\bf 60},
0231301 (1999).

\bibitem{books}
E.W. Kolb and M.S. Turner,  {\it The Early Universe}
(Addison-Wesley, Reading MA, 1990);\\
T. Padmanabhan, {\it Structure Formation in the Universe}
(Cambridge Univ. Press, Cambridge, 1993);\\
J.A. Peacock,  {\it Cosmological Physics} (Cambridge Univ.
Press, Cambridge, 1999);\\
A.R. Liddle and D.H. Lyth, {\it Cosmological Inflation and
Large-Scale Structure} (Cambridge Univ. Press, Cambridge, 2000).

\bibitem{ts}
C.G. Tsagas, Class. Quantum Grav. {\bf 19}, 3709 (2002)

\bibitem{Kulsrud}
R.M. Kulsrud, in {\em Galactic and Extragalactic Magnetic Fields}
Eds.  R. Beck, P.P. Kronberg and R. Wielebinski (Reidel, Dordrecht,
1990).

\bibitem{harmonics}
E.R. Harrison, Mod. Rev. Phys. {\bf 39}, 862 (1967);\\
M. Bruni, P.K.S. Dunsby, and G.F.R. Ellis, Ap. J. {\bf 395},
34 (1992);\\
P.K.S. Dunsby, M. Bruni, and G.F.R. Ellis, Ap. J. {\bf 395}, 54
(1992).

\bibitem{Marklund-Clarkson}
M. Marklund and C. Clarkson, Mon.\ Not.\ R.\ Astron.\ Soc. {\bf
    358}, 892 (2005).

\end{thebibliography}
\end{document}